\documentclass[10pt,notitlepage,pra,twocolumn]{revtex4}
\usepackage{amsfonts}
\usepackage{amsmath}
\usepackage{amssymb}
\usepackage{graphicx}
\usepackage[toc,page,header]{appendix}

\begin{document}

\title{Chirped Multi-photon adiabatic passage for a four-level ladder-type Rydberg excitation}
\author{Jing Qian$^\dagger$, Jingjing Zhai, Lu Zhang and Weiping Zhang}
\affiliation{Quantum Institute for Light and Atoms, Department of Physics, East China
Normal University, Shanghai 200062, People's Republic of China}

\begin{abstract}
We develop a multi-photon adiabatic passage to realize a highly efficient Rydberg excitation in a four-level ladder-type atomic system. The adiabatic passage is based on the existence of a novel quasi-dark state in the cascade excitation system where the frequencies of the lasers are appropriately chirped with time. We also investigate the influence of the interatomic Rydberg interaction on the passage and extend its application to the preparation of anti-blockade Rydberg atom pairs in the Rydberg blockade regime.
\end{abstract}

\maketitle
\preprint{}

\section{Introduction}

Recently, it has been recognized that quantum optical systems involving
highly-excited atomic Rydberg levels can provide a promising perspective for the applications in quantum information processing, quantum metrology, and quantum simulation \cite{Saffman10,Weimer10,Barato14,Peaudercerf14}. In particular, the huge polarizability of Rydberg states gives rise to giant Kerr coefficients which enable the nonlinear optical effects, such as electromagnetically induced transparency, in the single-photon level \cite{Baur14}. It also induces the strong dipole-dipole or van der Waals interactions between neighboring Rydberg atoms which can be used to implement atom-photon and photon-photon phase gates \cite{Lukin03,Friedler05,Petrosyan08}, effective photon-photon interactions \cite{Gorshkov11,He14}, and deterministic single-photon sources \cite{Shahmoon11,Honer11,Gorniaczyk14}.

In experimental realizations of these and other applications, it will be essential to have methods available to efficiently transfer ground-state atoms into and out of the highly-excited Rydberg states \cite{Deiglmayr06}. In the direct excitation case, except the requirement of deep ultraviolet wavelengths of the lasers, a poor selectivity of the excitation due to the optical mixing of the nearly degenerate Rydberg states limits its use \cite{Bajema04}. These shortcomings can be made up by a multi-step excitation via some intermediate states \cite{Brandenberger10,Johnson10,Ryabtsev11,Carr12,Carr13}. 
Especially, the stimulated Raman adiabatic passage (STIRAP) is a powerful technique that is used intensively in preparing and manipulating atomic and molecular states \cite{Bergmann98}. It is more robust to overcome the effects of the Doppler broadening, spatial laser-pulse inhomogeneities, and the spontaneous decay of the intermediate excited state. However, the ingenious formation of the dark state restricts the application of the STIRAP to the three-level excitation scheme. The multi-photon adiabatic passage and the fractional STIRAP with optimal pulse sequences are thus extensively studied \cite{Oreg84,Sola99,Gibson05,Maeda06,Topcu10,Maeda11} to realize highly efficient excitation in the four- or even more level schemes.

Compared to the general atomic excitation, strong interatomic interaction induced by the huge dipole moment of Rydberg states may result in poor effects of all above excitation strategies in the Rydberg excitation. The interaction shifts the original atomic energy levels and blocks the Rydberg excitation of the atoms whose neighboring atom has been excited to the Rydberg state, which is so-called Rydberg blockade effect \cite{Comparat10,Pritchard10,Viteau12}. To overcome this obstacle, one direct solution 
resorts to increasing the mean interatomic distance
(typically larger than $10\mu m$) by lowering atomic gas
density. Then the probability of obtaining Rydberg-excited atoms can
have a clear growth \cite{Beguin13}.
An alternative way is to utilize an appropriate optical detuning with respect to the Rydberg state to compensate the energy-shift induced by the Rydberg-Rydberg interaction, which can generate anti-blockade Rydberg atom pairs even in the deep Rydberg blockade regime \cite{Ates07,Amthor10,Qian13}. A modified STIRAP scheme based on this detuning-compensated approach is also proposed lately \cite{Yan11}.

In the present work, we design a multi-photon adiabatic passage to realize a highly efficient Rydberg excitation in a four-level ladder-type atomic system with two intermediate excited states located between the ground state and the Rydberg state. Differing from the previous works, the present adiabatic passage is based on a quasi-dark state which involves only one of the intermediate excited states.
This excitation scheme depends on the application of three coherent light pulses with carefully chosen Rabi frequencies, time widths, delay times, and most importantly, with chirped optical frequencies.   
The influence of the Rydberg-Rydberg interaction on this excitation scheme is investigated and discussed. We further find the detuning with respect to the other intermediate state could be an effective control knob to reduce the destructive effect of the Rydberg-Rydberg interaction to the excitation. Then the high excitation efficiency of the Rydberg atom pairs can be reached even in the deep Rydberg blockade regime where the adiabatic passage no longer exists. 

This paper is organized in the following way: In Sec. II,
we establish a detailed model for the quasi-dark state and multi-photon adiabatic passage in a four-level ladder-type atomic system without Rydberg-Rydberg interaction. The parameters of the optical pulses are optimized to make a trade off between the fidelity of the adiabatic passage based on the quasi-dark state and the spontaneous decay of the intermediate state involved. The influence of the Rydberg-Rydberg interaction on the excitation scheme is displayed and discussed with a two-atom model in Sec. III, while the crucial role for the optical detuning with respect to the other intermediate state in the strong interaction case is shown in Sec IV.

\section{Chirped Multi-photon Adiabatic Passage}

\begin{figure}
\includegraphics[width=3.34in,height=2.16in]{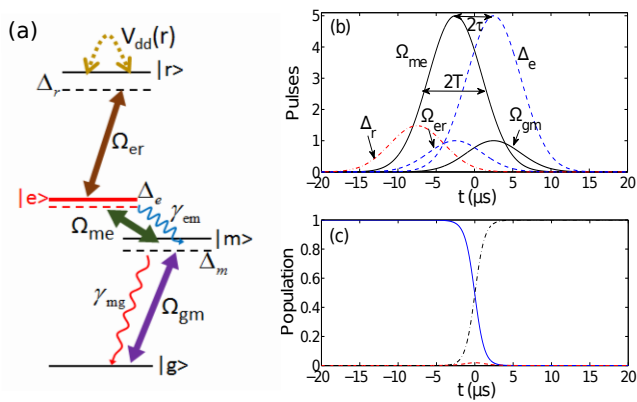}
\caption{(Color
online) (a) Scheme of the four-level ladder-type system interacting with the three laser fields with Rabi frequencies $\Omega _{gm}$, $\Omega _{me}$ and $%
\Omega _{er}$, and detunings $\Delta
_{m}$, $\Delta _{e}$ and $\Delta _{r}$ from the intermediate state $\left\vert m\right\rangle $, the intermediate state $\left\vert e\right\rangle $ and the Rydberg state $\left\vert r\right\rangle $, respectively. The two intermediate states are unstable and suffer from decay of rates $\gamma_{em}$ and $\gamma_{mg}$. $V_{dd}$ represents the interaction between two
neighboring Rydberg atoms. (b) Two pairs of pulse
sequences with counterintuitive character and their time-dependent Rabi
frequencies denoted by $\Omega _{gm}\left( t\right) $ and $\Omega
_{me}\left( t\right) $ (black solid curves), $\Delta _{e}\left( t\right) $
and $\Omega _{er}\left( t\right) $ (blue dashed curves), driving atoms from
ground state $\left\vert g\right\rangle $ to Rydberg state $\left\vert
r\right\rangle $ via a chirped multi-photon adiabatic passage. (c) The time evolution of the quasi-dark state $|D_4\rangle$ with the population probabilities of the component $|g\rangle$, $|r\rangle$, and $|e\rangle$ labeled by the blue solid curve, the black dash-dotted curve, and the red dashed curve, respectively. 
The population probability of $|e\rangle$ has been largely
suppressed by optimized pulses.}
\label{multi-level atom}
\end{figure}

As shown in Fig. \ref{multi-level atom}(a), we consider a four-level ladder-type atomic system with a ground state $|g\rangle$ and a Rydberg state $|r\rangle$, which can be excited via the two intermediate states $|m\rangle$ and $|e\rangle$. The transition Rabi frequencies between states $|i\rangle$ and $|j\rangle$ are denoted by $\Omega_{ij}$, with $ij=gm,me,er$, and optical detunings $\Delta_{m,e,r}$ may be applied with respect to the corresponding levels. All the Rabi frequencies are assumed to be real and positive in this work. The Rydberg and the two intermediate excited states decay by spontaneous emission of radiation, and we assume that the lifetime of the Rydberg state is much longer than the other two so that we can neglect its decay. In the present section, we will also neglect the dipole-dipole interaction $V_{dd}$ between the Rydberg atoms which may induce the Rydberg blockade effect, and focus on the single-atom excitation mechanism. Then in the rotating-wave frame the Hamiltonian
describing Rydberg excitation of such a single four-level atom is given by (where $\hbar =1
$ throughout the work)
\begin{eqnarray}
\mathcal{H}_{0} &=&\Delta _{m}\left\vert m\right\rangle \left\langle
m\right\vert +\Delta _{e}\left\vert e\right\rangle \left\langle e\right\vert
+\Delta _{r}\left\vert r\right\rangle \left\langle r\right\vert
\label{S_Ham} \\
&&+\left( \Omega _{gm}\left\vert m\right\rangle \left\langle g\right\vert
+\Omega _{me}\left\vert e\right\rangle \left\langle m\right\vert +\Omega
_{er}\left\vert r\right\rangle \left\langle e\right\vert +\text{H.c.}\right),
\notag
\end{eqnarray}
and the transition dynamics and the dissipations can be described by the master equation of single atom density matrix $\hat\rho$,
\begin{equation}
\partial _{t}\mathbf{\Hat{\rho}}\left( t\right) =-i\left[ \mathcal{H}%
_{0}\left( t\right) ,\mathbf{\Hat{\rho}}\left( t\right) \right] +\mathcal{L}%
\left[ \mathbf{\Hat{\rho}}\left( t\right) \right],   \label{Master_single}
\end{equation}%
where the Lindblad operator $\mathcal{L}\left[ \mathbf{\Hat{\rho}}\left(
t\right) \right] $ referring to the spontaneous decays from the two intermediate
unstable states $\left\vert m\right\rangle $ and $\left\vert e\right\rangle $
is readily written as%
\begin{eqnarray}
\mathcal{L}\left[ \mathbf{\Hat{\rho}}\right]  &=&\frac{\gamma _{mg}}{2}%
\left( 2\hat{\sigma}_{gm}\mathbf{\Hat{\rho}}\Hat{\sigma}_{gm}^{\dagger
}-\left\{ \Hat{\sigma}_{mm},\mathbf{\Hat{\rho}}\right\} \right) 
\label{Lindlabd_Sin} \\
&&+\frac{\gamma _{em}}{2}\left( 2\hat{\sigma}_{me}\mathbf{\Hat{\rho}}\hat{%
\sigma}_{me}^{\dagger }-\left\{ \hat{\sigma}_{ee},\mathbf{\Hat{\rho}}%
\right\} \right)   \notag
\end{eqnarray}%
with $\hat{\sigma}_{ij }=\left\vert i \right\rangle\left\langle j \right\vert $ denoting the atomic transition operator and $\gamma _{em\left( mg\right) }$ the spontaneous decay rates with respect to states $|e\rangle$ $ (|m\rangle) $. An efficient population transfer from the ground state $|g\rangle$ to the Rydberg state $|r\rangle$ can be relatively easily achieved in experiments via a multiple photon cascade excitation, with dissipation from the intermediate states a major impediment \cite{Qian14}. 

Recently the technique of STIRAP is employed widely in the Rydberg excitation of three-level ladder atomic systems due to its advantage of immunity to the decay of the intermediate state \cite{Cubel05,Petrosyan13}. This feature arises from
the existence of a ``dark'' eigen-state $\left\vert D_{3}\right\rangle $ in a coherent coupled three-level system, with eigen-energy $\lambda _{3}=0$ and eigen-vector%
\begin{equation}
\left\vert D_{3}\right\rangle =\cos \theta \left\vert g\right\rangle -\sin
\theta \left\vert r\right\rangle,   \label{darkSta}
\end{equation}%
where the mixing angle $\theta$ is given by $\tan \theta =\Omega_{g}/\Omega_{r}$ with $\Omega_{g(r)}$ the Rabi frequency between the intermediate state and the state $|g\rangle$ $(|r\rangle)$. It does not have any contribution from the unstable intermediate state. Then a complete transfer from $|g\rangle$ to $|r\rangle$ can be achieved by adiabatically changing the dark state superposition, that is, performing a pair of optical pulses in a counterintuitive order \cite{Bergmann98,Ditter07}. 

For a four-level atomic system considered in the present work, 
by diagonalizng Hamiltonian (\ref{S_Ham}) we find due to the different symmetry from the three-level system, there is no dark state which is immune to the two intermediate states $|m\rangle$ and $|e\rangle$. However, as long as the condition 
\begin{equation}
\Omega _{er}^{2}=\Delta_{e}\Delta_{r} \label{condition}
\end{equation}
is fulfilled, there exists a novel eigen-state $|D_4\rangle$ with its eigen-energy $\lambda _{4}=0$ and eigen-vector
\begin{equation}
\left\vert D_{4}\right\rangle =\cos \varphi \left\vert g\right\rangle +\sin
\varphi \cos \phi \left\vert e\right\rangle -\sin \varphi \sin \phi
\left\vert r\right\rangle,   \label{quasiDark}
\end{equation}%
which involves one of the intermediate states, $|e\rangle$, and is so called quasi-dark state by us. Comparing to state $\left\vert
D_{3}\right\rangle $, $\left\vert D_{4}\right\rangle $ possesses two mixing
angles which are given by
\begin{eqnarray}
 \tan \varphi &=&-\frac{ \Omega _{gm}}{\Omega _{me}} \sqrt{1+\frac{\Delta _{e}^{2}}{\Omega _{er}^{2}}},\\
 \tan \phi &=&\frac{\Delta_{e}}{\Omega _{er}}. 
\end{eqnarray}
Similar to the three-level case, one can build a multi-photon adiabatic passage via state $\left\vert D_{4}\right\rangle $. The atom initially preparing in the ground state $\left\vert g\right\rangle $ ($\langle g | D_{4}\rangle =1$ for $\cos \varphi =1$ and $\cos \phi =0$) will finally settle in the Rydberg state $\left\vert r\right\rangle $ ($\langle r |D_{4}\rangle =1$ for $\sin \varphi =1$ and $\sin \phi =1$) by slowly adjusting the two mixing angles. It is a bit special that except the three Rabi frequencies $\Omega _{gm}$, $\Omega _{me}$, and $\Omega _{er}$, the detunings $\Delta_e$ and $\Delta_r$ should also change over time to meet the quasi-dark condition (\ref{condition}). This could be achieved through coupling state $|m\rangle$ and state $|e\rangle$, state $|e\rangle$ and state $|r\rangle$ by two frequency-chirped optical pulses, the technology of which has been widely investigated in the theory and the experiment of atomic and molecular state control \cite{Malinovsky01,Conover02,Beterov11,Prozument11,Torosov11,Maeda211}. 
The time sequence of pulses and the time dependence of detunings for our chirped multi-photon adiabatic passage are displayed in Fig. \ref{multi-level atom}(b).  

\begin{figure}
\includegraphics[width=2.5in]{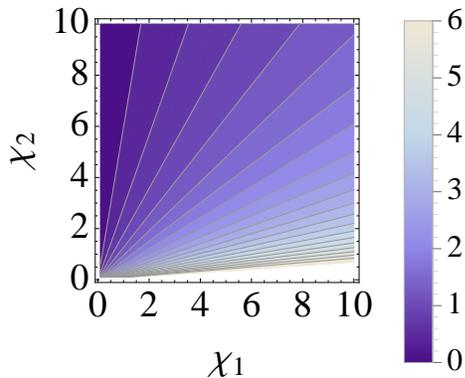}
\caption{(Color online) The accumulated population probability on the excited state $\left\vert e\right\rangle $ is estimated by the characteristic function $f\left( \protect\chi _{1}, \protect\chi _{2}\right) $. Areas marked by darker color mean smaller values of $f$.}
\label{int}
\end{figure}

Since the quasi-dark state $\left\vert D_{4}\right\rangle $ involves intermediate state $|e\rangle$ of non-ignorable decay, which may break the adiabatic transfer, we first calculate the accumulated population probability on the component state $\left\vert e\right\rangle $ of state $\left\vert D_{4}\right\rangle $ over the whole transfer process, whose value can briefly estimate that destructive effect. For generality, we assume all the pulses and chirped detunings are Gaussian, with the same width $T$ and the same duration $\tau$,
\begin{eqnarray}
\Omega _{gm}\left( t\right)  &=&\Omega _{gm}^{\max }e^{-\left( t-\tau
\right) ^{2}/T^{2}},\notag\\
\Omega _{me}\left( t\right) &=&\Omega _{me}^{\max
}e^{-\left( t+\tau \right) ^{2}/T^{2}}  \label{Pulses} \notag\\
\Delta _{e}\left( t\right)  &=&\Delta _{e}^{\max }e^{-\left( t-\tau \right)
^{2}/T^{2}},\\
\Omega _{er}\left( t\right) &=&\Omega _{er}^{\max }e^{-\left(
t+\tau \right) ^{2}/T^{2}} ,\notag \\
\Delta _{r}\left( t\right)  &=&\frac{(\Omega _{er}^{\max })^2}{\Delta_e^{max}}e^{8\tau^{2}/T^{2}}e^{-\left( t+3\tau \right) ^{2}/T^{2}},\notag
\end{eqnarray}
where the quantities with superscript ``max'' represent the Gaussian peak values. $\tau >0$ is required to ensure the counterintuitive feature of STIRAP pulses.
Then according to the equation (\ref{quasiDark}), the time evolution of the population probabilities on the three components of quasi-dark state $|D_4\rangle$ is plotted in Fig. \ref{multi-level atom}(c),
and the accumulated population probability on the component state $\left\vert e\right\rangle $ is given by
\begin{equation}
\int\limits_{-\infty }^{+\infty }\left\vert \sin \varphi \cos \phi
\right\vert ^{2}dt=\frac{T^{2}}{8\tau }f\left( \chi _{1},\chi _{2}\right), 
\label{popu_e}
\end{equation}%
where we have defined 
\begin{equation}
f\left( \chi _{1},\chi _{2}\right) =\int_{-\infty }^{+\infty }d\xi 
\frac{\chi _{1}^{2}}{\chi _{1}^{2}\chi _{2}^{2}e^{\xi }+e^{-\xi }+\chi
_{1}^{2}},
\end{equation}
as a function of the two peak value ratios $\chi _{1}=\Omega _{gm}^{\max }/\Omega _{me}^{\max }$ and $%
\chi _{2}=\Delta _{e}^{\max }/\Omega _{er}^{\max }$. We plot $f(\chi_1,\chi_2)$ in Fig. \ref{int} which clearly shows that the condition $\chi _{2}>\chi _{1}$ is preferred for a small accumulated population probability and thus a weak dissipation effect. This is due to the condition gives $\Delta
_{e}^{\max }>\Omega _{gm}^{\max }\Omega _{er}^{\max }/\Omega _{me}^{\max }$ which means a far-off resonant excitation to the intermediate state $|e\rangle$.
According to Eq. (\ref{popu_e}), it seems that a short pulse width and a long pulse duration can also reduce the accumulated population probability. However, 
it means a rapidly changed pulse sequence which goes against the adiabatic passage theory \cite{Qian10}.

\begin{figure}
\includegraphics[width=3.49in,height=1.54in]{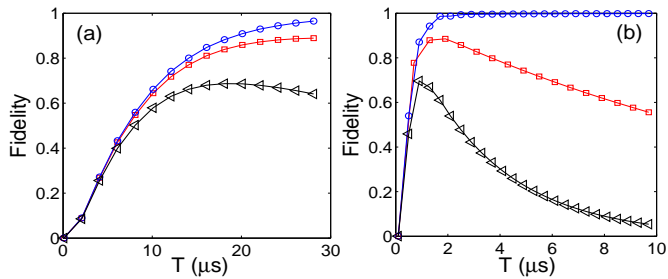}
\caption{(Color online) Fidelity $%
F^{\left( \infty \right) }$ as a funtion of pulse width $T$ for different
spontaneous decay rates of state $\left\vert e\right\rangle $%
: $\protect\gamma _{em }=0$ (blue curve with circles), $0.1$%
MHz (red curve with squares), $0.5$MHz (black curve with triangles). (a) $%
\Omega _{gm}^{\max }=\Omega _{er}^{\max }=1.0$MHz, $\Omega _{me}^{\max
}=\Delta _{e}^{\max }=5.0$MHz and (b) $\Omega _{gm}^{\max }=\Omega
_{er}^{\max }=\Omega _{me}^{\max }=\Delta _{e}^{\max }=5.0$MHz. In
simulations other parameters are $\gamma_{mg}=0.1$ MHz, $\protect\tau =0.5T$, $\Delta _{m}=0$, $%
\Delta _{r}=\Omega _{er}^{2}/\Delta _{e}$.}
\label{Fidelity}
\end{figure}

To investigate the influence of these relevant parameters on the chirped multi-photon adiabatic passage more precisely, we define the fidelity of the quasi-dark state $F^{\left( t\right) }=\left\langle D_{4}\left( t\right)
\right\vert \mathbf{\Hat{\rho}}\left( t\right) \left\vert D_{4}\left(
t\right) \right\rangle $ and simulate the master equation (\ref{Master_single}) to obtain its value at $t\rightarrow\infty $ for various parameters. Notice that $F^{\left(\infty \right) }$ is just the population of the Rydberg state $|r\rangle$. 
As shown in Fig. \ref{Fidelity},
when there is no dissipation of the intermediate state $|e\rangle$ ($\gamma_{em}=0$, which means $%
\left\vert D_{4}\left( t\right) \right\rangle $ is a perfect dark state), the blue curves labeled by circles show the value of $F^{\left(
\infty \right) }$ can reach unity as long as the applied pulse for
optical excitation is sufficiently long so that the dark state can be adiabatically followed. In the presence of nonzero $\gamma_{em}$, we compare the case that $\Omega _{gm}^{\max }=\Omega _{er}^{\max }=1.0$MHz and $\Omega _{me}^{\max }=\Delta _{e}^{\max }=5.0$MHz (Fig. \ref{Fidelity}(a)) which meets the decay-suppressed condition $\chi_2>\chi_1$ with the case that $\Omega _{gm}^{\max }=\Omega _{er}^{\max
}=\Omega _{me}^{\max }=\Delta _{e}^{\max }=5.0$MHz (Fig. \ref{Fidelity}(b)) which does not meet the condition. In both cases, the fidelity decreases as the decay rate $\gamma_{em}$ increases. As the pulse width $T$ grows, there is an obvious single-peak structure of the fidelity appears in the second case, which is different from the typical STIRAP case.
That is due to our chirped multi-photon adiabatic passage (CMPAP) is based on a quasi-dark state. For short pulses the adiabatic following can not be guaranteed while for long pulses the influence of the decay of state $|e\rangle$ becomes significant, which will also breaks the adiabaticity.
In a real experimental realization, the rubidium atom system as used in Rydberg excitation experiments \cite{Ryabtsev11} and \cite{Thoumany09} is a suitable candidate to adopt our approach. The selected four levels of rubidium atoms in those experiments can build a ladder-type atomic system required by the CMPAP, e.g., $5S_{1/2}(|g\rangle)$, $5P_{3/2}(|m\rangle)$,  $6S_{1/2}$ or $5D_{5/2}$($|e\rangle$), and $63P_{3/2}(|r\rangle)$. The natural linewidths of the two intermediate states are in the order of several MHz that is consistent with the values applied in our numerical simulations.

By then we conclude that without the Rydberg-Rydberg
interaction it is possible to achieve an easy and high-fidelity CMPAP to
coherent transfer ground-state atoms into the high-lying Rydberg state in a four-level ladder system.

\section{Interacting Rydberg atoms}

With two or more atoms the Rydberg blockade effect comes into play, which will give rise to an energy shift on the single atom Rydberg state. To investigate its influence to the performance of the CMPAP, we consider a two-atom model whose Hamiltonian is written as
\begin{equation}
\mathcal{H}_{I}=\mathcal{H}_{0}\otimes \mathcal{I}+\mathcal{I}\otimes 
\mathcal{H}_{0}+V_{dd}\left\vert rr\right\rangle \left\langle rr\right\vert ,
\label{Ham}
\end{equation}
where $\mathcal{H}_{0}$ is the four-level single-atom Hamiltonian given by Eq. (\ref{S_Ham}), $\mathcal{I}$ is the identity matrix, and $V_{dd}$ represents the dipole-dipole interaction strength between two atoms both on the Rydberg state. The corresponding master equation becomes
\begin{equation}
\partial _{t}\mathbf{\Hat{\rho}_I}\left( t\right) =-i\left[ \mathcal{H}%
_{I}\left( t\right) ,\mathbf{\Hat{\rho}_I}\left( t\right) \right] +\sum_{j=1,2}\mathcal{L}_j%
\left[ \mathbf{\Hat{\rho}_I}\left( t\right) \right],   \label{Master_two}
\end{equation}%
where the form of the Lindblad operator $\mathcal{L}_j$ is same as Eq. (\ref{Lindlabd_Sin}) but with the atomic transition operators for atom $j$ only.
The subscript ``$I$'' indicates the case for two interacting atoms.

The eigenstates of $\mathcal{H}_{I}$ may be quite different to the single atom case due to the Rydberg interaction term so as to threaten the existence of the quasi-dark state $|D_4\rangle$. 
In the weak interaction case where $%
|V_{dd}|\ll \Omega$ with single-atom characteristic frequency $\Omega\in  
\{ \Omega
_{gm}^{\max },\Omega _{me}^{\max },\Omega _{mr}^{\max },\Delta _{e}^{\max } \} $, the CMPAP for Rydberg excitation of both atoms are sustained with the two-atom eigen-state which can be approximately represented as $\left\vert D_{4}\right\rangle_I =\left\vert
D_{4}\right\rangle \otimes \left\vert D_{4}\right\rangle $. Then there still exists an adiabatic passage between the ground state $|gg\rangle$ and the double Rydberg excited state $|rr\rangle$. However, this will not be the case for $%
|V_{dd}|\sim \Omega$ or even $|V_{dd}|>\Omega$, where the target state for the CMPAP could be no longer the state $|rr\rangle$ due to the Rydberg blockade effect \cite{Petrosyan13}.

A similar example of this is in the three-level Rydberg excitation system. When the detuning with respect to the intermediate state is zero, the single-atom dark state $|D_3\rangle$ is replaced by a new two-atom quasi-dark state \cite{Moller08},
\begin{eqnarray}
\left\vert D_{3}\right\rangle_I  &=&\frac{1}{\sqrt{\cos ^{4}\theta +2\sin
^{4}\theta }}\left[ \left( \cos ^{2}\theta -\sin ^{2}\theta \right)
\left\vert gg\right\rangle \right.   \label{darkTwo} \\
&&\left. -\cos \theta \sin \theta \left( \left\vert gr\right\rangle
+\left\vert rg\right\rangle \right) +\sin ^{2}\theta \left\vert
mm\right\rangle \right] ,  \notag
\end{eqnarray}%
which means while the STIRAP process in the single atom case ensures that the Rydberg state $|r\rangle$ is finally populated, due to the Rydberg-Rydberg interaction the pair of atoms will be adiabatically steered into a superposition state of the ground state $|gg\rangle$ and the intermediate state $|mm\rangle$.
The situation can be modified by applying a finite and appropriate detuning from 
the intermediate state, which might generate a new eigen-state that involves both $\left\vert gg\right\rangle $ and $\left\vert rr\right\rangle $, leading
to a near-unity final population of $\left\vert rr\right\rangle $ \cite{Rao14}. 
However, since in the case of our four-level Rydberg atoms excited by the CMPAP, the detuning $\Delta_m$ and $\Delta_e$ are both time-dependent, the same approach can not be adopted directly. We then first perform a numerical simulation on the excitation dynamics by the two-atom master equation (\ref{Master_two}) to investigate the influence of the interaction strength $V_{dd}$ on the two-atom eigen-states and on the CMPAP. 

\begin{figure}
\includegraphics[width=3.27in,height=2.69in]{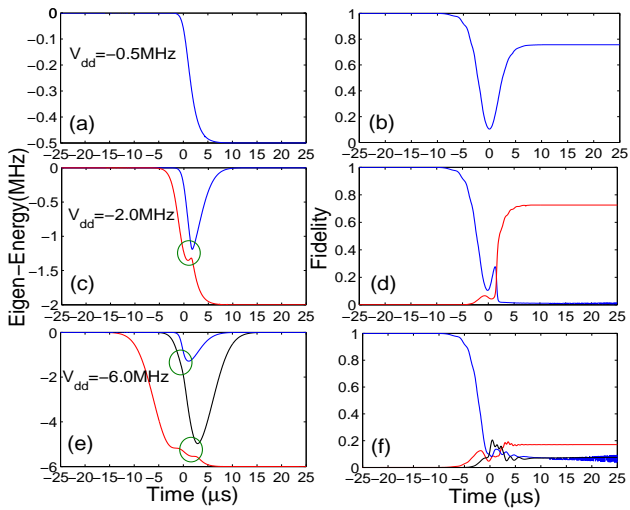}
\caption{(Color online) The left column:
Time evolution of the energies of the eigen-states involved in the transition from $\left\vert gg\right\rangle$
to $ \left\vert rr\right\rangle $ for the cases, from top to bottom,  $V_{dd}=-0.5$ MHz, $%
-2.0$ MHz and $-6.0$ MHz; The right column: the
corresponding fidelity (population) in these eigen-states. Other parameters are
same as Fig. \protect\ref{Fidelity}(b), note that $\Delta _{m}=0$.}
\label{EnergyFidelity}
\end{figure}

We focus on the several two-atom eigen-states $|E_j\rangle$ which can serve as the double Rydberg excitation passages, that is, $|E_j(t=-\infty)\rangle=|gg\rangle$ and $|E_j(t=+\infty)\rangle=|rr\rangle$ when the same pulse sequences and the chirped detunings as discussed in the last section are applied. The time evolutions of their eigen-energys and fidelities $%
F_{j}^{\left( t\right) }=\left\langle E_{j}\left( t\right) \right\vert 
\mathbf{\Hat{\rho}}_{I}\left( t\right) \left\vert E_{j}\left( t\right)
\right\rangle $ under the different Rydberg interaction strengths are displayed in Fig. \ref{EnergyFidelity}. In the weak
interaction case ($\left\vert V_{dd}\right\vert \ll \Omega $) we find there exists a single two-atom eigen-state which can connect $|gg\rangle$ and $|rr\rangle$. However, its eigen-energy changes over time so that it is not a dark or even quasi-dark state. The fidelity of this state reveals a deep at $t=0$ at
which the effective optical coupling approaches its maximal value, after
then the fidelity quickly revives and attains as high as $0.76$ in the region where the eigen-state turns into $|rr\rangle$, which means a large population on the double Rydberg excited state. In the intermediate case ($\left\vert V_{dd}\right\vert \sim \Omega$) we still find a large final population probability on $|rr\rangle$ but it is not due to an adiabatic passage. In this case there is not one single two-atom eigen-state which connects $|gg\rangle$ and $|rr\rangle$ as time going, but are two different states $|E_g(t)\rangle$ (blue curve) and $|E_r(t)\rangle$ (red curve) which are initially $|gg\rangle$ and finally $|rr\rangle$, respectively. Moreover, as marked by a green cycle in the figure, at the region near $t=0$ there is an avoided crossing with a small energy gap between these two states, which may result in an efficient population transfer into the final state $|rr\rangle$. 
In the strong interaction case ($\left\vert V_{dd}\right\vert \gg \Omega$), $|E_g(t)\rangle$ and $|E_r(t)\rangle$ can no longer be connected directly by an avoided crossing. It instead requires another eigen-state to play the role as the bridge (see the black curve) between them. As the interaction strength $|V_{dd}|$ increases, more and more eigen-states are involved in the population transfer dynamics, which scatters the population and finally results in a poor final population on $|rr\rangle$.

Our simulation shows that the CMPAP are very effective in the Rydberg excitation with a weak Rydberg-Rydberg interaction. When the interaction is comparable with the characteristic energy of the single atom the adiabatic passage will be taken apart, but due to the small energy gap between the parts the Rydberg excitation efficiency of the CMPAP is still pretty robust. Finally the CMPAP will fail in an atomic system with dominant Rydberg-Rydberg interaction.

\section{Intermediate-state detunings}

\begin{figure}
\includegraphics[width=2.7in]{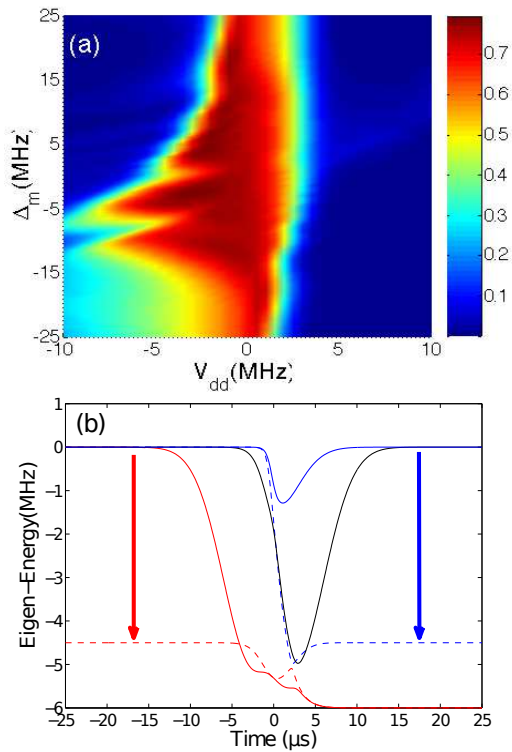}
\caption{(Color online) (a) Fidelity $%
F_{rr}^{\infty }$ standing for the probability of anti-blockade atom pair
production in the parameter space of $V_{dd}$ and $\Delta _{m}$ with other parameters $\Omega _{me}^{\max }=\Omega
_{gm}^{\max }=\Omega _{er}^{\max }=\Delta _{e}^{\max }=5.0$ MHz and $T=5%
\protect\mu $s, $\protect\gamma _{em}=\protect\gamma _{mg}=0.1$MHz. 
(b) The time evolution of the energy of the eigen-states involved in the excitation. The solid curves are for the case $\Delta _{m}=0$ and the dashed curves are for $\Delta _{m}=-4.5$ MHz. The Rydberg-Rydberg coupling strength $\left\vert
V_{dd}\right\vert $($=6$MHz) is larger than the characteristic frequency $\Omega $($=5$MHz)%
. Eigen-states $\left\vert E_{g}\right\rangle $, $\left\vert
E_{m}\right\rangle $ (assistant eigen-state) and $\left\vert
E_{r}\right\rangle $ are marked with the blue, the black and the red curves, respecttively. At time $%
t=-25\protect\mu $s, $\left\vert E_{r}\right\rangle \approx \left(
\left\vert me\right\rangle -\left\vert em\right\rangle \right) /\protect%
\sqrt{2}$; at $t=25\protect\mu $s, $\left\vert E_{g}\right\rangle \approx
\left( \left\vert gm\right\rangle -\left\vert mg\right\rangle \right) /%
\protect\sqrt{2}$.}
\label{double_exct}
\end{figure}

Recent researches illustrate that the optical detuning is an effective control knob in the Rydberg excitation mechanism. Typical examples include the anti-blockade excitation with the detuning to the Rydberg state compensating the frequency shift caused by the Rydberg-Rydberg interaction \cite{Ates07,Amthor10}, and a new two-atom adiabatic passage in the three-level atomic system with an appropriate detuning to the intermediate state as proposed by Rao and M\o lmer \cite{Rao14}.
In our four-level excitation scheme, although the detuning to the Rydberg state $|r\rangle$ and the detuning to the intermediate state $|e\rangle$ are both limited by the quasi-dark sate condition (\ref{condition}), the detuning to the other intermediate state $|m\rangle$ is a good candidate to optimize the effect of our CMPAP in the strong interaction case.

In Fig. \ref{double_exct}(a) we show the fidelity $F_{rr}^{\infty }=\left\langle
rr\right\vert \mathbf{\Hat{\rho}}_{I}\left\vert rr\right\rangle $, i.e., the final probability in state $\left\vert
rr\right\rangle $ as a function of the detuning $\Delta _{m}$ and the interaction $V_{dd}$ under the same parameters as used in Fig. \ref{EnergyFidelity}. 
When $V_{dd}=0$ it shows that the value of the fidelity is high and has less dependence on $\Delta_m$. This can be attributed to the adiabatic passage based on the two-atom quasi-dark state $|D_{4}\rangle_I $ which does not involve the intermediate state $|m\rangle$.
As the interaction $V_{dd}$ becomes remarkable, things have certainly changed. High fidelity disappears in the most of the area but still survives in lower left corner of the figure. For example when $V_{dd}=-6.0$MHz and $%
\Delta _{m}=-4.5$MHz, $F_{rr}^{\infty }\approx 0.71$. Notice that when $\Delta_m=0$ under the same value of the interaction $V_{dd}$ the fidelity is very poor ($F_{rr}^{\infty }\approx 0.17$), the cause of which has been analyzed in Fig. \ref{EnergyFidelity}(e) and 4(f) that the original adiabatic passage is split and other eigen-states are involved into the excitation. Since in the strong interaction case the adiabatic passage is no longer based on the quasi-dark state, we attribute the improvement of the fidelity to the change of the two-atom eigen-states and their eigen-energies with a nonzero $\Delta_m$. 
To display the change, in Fig. \ref{double_exct}(b) we plot the time evolution of the eigen-energy spectrum of the involved eigen-states under the different values of $\Delta_m$. The solid curves are for the case $\Delta_m=0$ which is same as Fig. \ref{EnergyFidelity}(e) while the dashed curves are for the case $\Delta _{m}=-4.5$ MHz. We find the eigen-state $|E_g(t)\rangle$ (blue curves) is initially $|gg\rangle$ and in the end almost equals to $(|gm\rangle-|mg\rangle)/\sqrt{2}$ while the eigen-state $|E_r(t)\rangle$ (red curves) changes from $(|em\rangle-|me\rangle)/\sqrt{2}$ to $|rr\rangle$ as time going. For the zero detuning case in the optical excitation region (about $t=0$) there is a large energy gap between these two states and another eigen-state is involved to fill the passage to the double Rydberg excitation which results in the final low fidelity. Since both the evolution of state $|E_g(t)\rangle$ and state $|E_r(t)\rangle$ involve the intermediate state $|m\rangle$, an appropriate value of detuning $\Delta_m$ may reduce the energy gap between them as identified by the dashed curves, and then the CMPAP recovers a high Rydberg excitation rate.

\section{Conclusion}

In summary, we present a highly efficient three-photon cascade excitation scheme for Rydberg atoms. Different from the typical two-photon excitation scheme with the STIRAP, our scheme is based on a novel quasi-dark eigen-state which appears when the frequencies of the optical pulses are appropriately chirped. It enables a CMPAP between the ground state and the Rydberg state in the four-level atomic system. When the interatomic Rydberg interaction is considered, we find the two-atom quasi-dark state approach is only feasible when the interaction strength is relatively weak compared to the peak value of the optical Rabi frequencies. When they two become comparable or even the Rydberg interaction is dominant, the adiabatic excitation passage to the Rydberg atom pairs will be broken. Moreover, we find an appropriate optical detuning with respect to the intermediate state can reduce the energy gap between the fragment states of the broken adiabatic excitation passage. With its help the CMPAP give rise to an enhancement of the production of anti-blockade Rydberg atom pairs even in the strong Rydberg blockade regime.

This work was supported by the National Basic Research Program of China (973 Program) under Grant No.2011CB921604, the NSFC under Grants No. 11474094, No. 11104076 and No. 11234003, and the
Specialized Research Fund for the Doctoral Program of Higher Education No.
20110076120004.

\end{document}